\pdfoutput=1

\documentclass[3p,times,twocolumn]{elsarticle}

\usepackage{ecrc}

\volume{00}

\firstpage{1}

\journalname{Nuclear Physics B Proceedings Supplement}

\runauth{}

\jid{nuphbp}

\jnltitlelogo{Nuclear Physics B Proceedings Supplement}

\usepackage{amssymb}

\usepackage[figuresright]{rotating}

\begin{document}

\begin{frontmatter}



\dochead{}

\title{Origin of Galactic Cosmic Rays}


\author{Pasquale Blasi}

\address{INAF/Osservatorio Astrofisico di Arcetri\\
Largo E. Fermi, 5 - 50125 Firenze, Italy}

\begin{abstract}
The origin of the bulk of cosmic rays (CRs) observed at Earth is the topic of a century long investigation, paved with successes and failures. From the energetic point of view, supernova remnants (SNRs) remain the most plausible sources of CRs up to rigidity $\sim 10^{6}-10^{7}$ GV. This confidence somehow resulted in the construction of a paradigm, the so-called SNR paradigm: CRs are accelerated through diffusive shock acceleration in SNRs and propagate diffusively in the Galaxy in an energy dependent way. Qualitative confirmation of the SNR acceleration scenario has recently been provided by gamma ray and X-ray observations. Diffusive propagation in the Galaxy is probed observationally through measurement of the secondary to primary nuclei flux ratios (such as B/C). 
There are however some weak points in the paradigm, which suggest that we are probably missing some physical ingredients in our models. The theory of diffusive shock acceleration at SNR shocks predicts spectra of accelerated particles which are systematically too hard compared with the ones inferred from gamma ray observations. Moreover, hard injection spectra indirectly imply a steep energy dependence of the diffusion coefficient in the Galaxy, which in turn leads to anisotropy larger than the observed one. Moreover recent measurements of the flux of nuclei suggest that the spectra have a break at rigidity $\sim 200$ GV, which does not sit well with the common wisdom in acceleration and propagation. In this paper I will review these new developments and suggest some possible implications.  
\end{abstract}


\end{frontmatter}


\section{Introduction}
\label{sec:intro}

Understanding the origin of cosmic rays means to be able to unfold a complex chain of physical processes that go from the acceleration of a few charged particles in potentially unknown sources to their propagation in the interstellar medium (ISM). Each of these steps involves plasma physics processes often in a non linear regime. The whole picture is usually made fuzzier by the poor knowledge of the values of environmental parameters (e.g. densities, temperatures, fraction of ionized material). Moreover most observations involve quantities (such as fluxes, chemical composition, anisotropy) measured at Earth that are averages over long (propagation) times and over numerous sources, potentially very different. It is not surprising that a century after the discovery of cosmic rays we are still debating about some aspects of this problem. The purpose of this short review is to provide a biased view of some aspects of the theory of the origin of CRs that I think are understood and some that are not understood or are poorly understood. I will limit myself to CRs that we think originate inside the Galaxy, though the definition of this transition energy is somehow part of the aspects that need to be discussed. 

So far, the only theoretical framework that reached a sufficient level of elaboration to deserve the name ``model'' is the one based on SNRs as the main sources of the bulk of Galactic CRs. In fact, the lack of a reasonable alternative has elevated the model to the rank of a paradigm. CR acceleration is believed to occur at the forward shock of SNRs through diffusive shock acceleration (DSA). The test particle theory of DSA (see \cite{blandford} for a review) predicts that the spectrum of accelerated particles is $N(E)\propto E^{-\gamma}$ with $\gamma=(r+2)/(r-1)$, where $r$ is the compression factor of the shock. For typical parameters of SNR shocks, $r\approx 4$ and $\gamma=2$. The propagation of CRs in the Galaxy is usually parametrized through a diffusion coefficient $D(E)\propto E^{\delta}$. For nuclei for which spallation is negligible the equilibrium spectrum observed at Earth is $n(E)\propto N(E)/\tau_{esc}(E)$, where $\tau_{esc}(E)\sim H^{2}/D(E)\sim E^{-\delta}$, where $H$ is the size of the halo of the Galaxy. It follows that $n(E)\propto E^{-\gamma-\delta}$. Comparison with observations leads to $\delta\approx 0.7$ if $\gamma\approx 2$. $\delta\approx 0.6-0.7$ is also compatible with the low energy observed slope of the B/C ratio. This might look like a self-consistent picture at first sight, but two problems immediately arise: 1) the observed gamma ray spectra in the $1-100$ GeV range suggest a steeper CR spectrum, with $\gamma\sim 2.3-2.4$ \cite{dam}; 2) A diffusion coefficient $D(E)\propto E^{0.7}$ would lead to a large scale CR anisotropy much larger than observed \cite{amato2}. 
The non-linear theory of DSA (NLDSA, see \cite{malkovdrury} for a review) makes these two problems even more severe: the theory accounts for the dynamical reaction of accelerated particles on the shock, which is responsible for the formation of a precursor upstream of the shock. In NLDSA the compression factor felt by accelerated particles becomes a function of energy, which reflects in concave spectra, which turn out to be even harder than $E^{-2}$ at $E>10-100$ GeV. As shown in \cite{berevolk07}, injection of CRs in SNRs through NLDSA leads to require $D(E)\propto E^{0.75}$, that is hardly compatible with the observed large scale anisotropy. 

It is clear that the problems of acceleration in the sources and propagation in the ISM are tightly connected with each other and cannot be studied as two independent problems. For instance, the assumption that the diffusion coefficient $D(E)$ is a given function of energy is most likely inappropriate, in that CRs themselves are able to create the scattering centers responsible for their diffusive motion. In turn, the diffusion coefficient affects the CR spectrum observed at Earth. This phenomenon, discussed by \cite{plesser,serpico} might play an important role in understanding features on the CR spectrum recently reported by PAMELA \cite{pam} and CREAM \cite{cream}. 

This paper is organized as follows: in \S \ref{sec:nldsa} I will discuss the implications of the NLDSA in terms of magnetic field amplification and how this might mitigate the spectral problem illustrated above. In \S \ref{sec:propa} I will discuss the important role of CRs to self-generating the scattering centers on which diffusion in the ISM takes place. The implications for nuclear spectra will also be illustrated. In \S \ref{sec:neutrals} I will discuss an important recent development in the theory of CR acceleration in SNRs, in connection with the presence of neutrals hydrogen in the acceleration region. In \S \ref{sec:trans} I will comment on the implications of the SNR paradigm for the transition from Galactic to extragalactic CR. A summary will be presented in \S \ref{sec:summary}.

\section{Magnetic field amplification, Maximum energy and velocity of scattering centers}
\label{sec:nldsa}

NLDSA describes the process of particle acceleration at collisionless shocks relaxing the {\it test particle} assumption, namely taking into account the CR dynamical reaction on the shock. Moreoever, \cite{amato2,damLett,damLong} also introduced in the theory the phenomenon of CR-induced magnetic field amplification and the dynamical reaction of the magnetic  field on the background plasma. 

The magnetic field can be amplified because of the streaming instability induced by the super-alfvenic motion of CRs in the upstream plasma \cite{bell78,achterberg,zweibel}. The instability leads to resonant growth of modes with wavenumber $k\sim 1/r_{L}$, where $r_{L}$ is the Larmor radius of particles that dominate the particle number at the shock. The waves produced by accelerated particles can also resonantly be absorbed by the same particles thereby leading to their diffusive motion. More recently \cite{bell2004} found that non resonant modes may grow much faster than the resonant modes on scales $k\gg 1/r_{L}$. Although this phenomenon may be relevant for magnetic field amplification, it is probably not the main mechanism for scattering the particles and for their acceleration at the shock.

The amplification of magnetic field due to the streaming of CRs with the shock was first introduced in the theory of NLDSA by \cite{amato2006,ellison}. Its effect for achieving higher maximum energies was later investigated in \cite{dam2007}. Magnetic field amplification is also needed to explain the narrow X-ray rims detected in virtually all young SNRs (see \cite{rims} and references therein for a recent review). 

Finally the creation of turbulent magnetic field in the shock region might have an important effect in determining the shape of the spectrum. As first emphasized by \cite{bell78}, the relevant shock compression factor for particle acceleration is the ratio of velocity of the scattering centers. If the waves are slow in the frame comoving with the plasma, then this is basically the same as the compression factor of plasma speeds. On the other hand, in the non-linear regime, the waves might acquire a substantially higher speed (perhaps of the order of the Alfv\'en speed calculated in the amplified field). In this case the relevant compression factor may be somewhat smaller and reflect into steeper spectra of accelerated particles \cite{damiano,ptuskin2010}. It is worth recalling that the details of this phenomenon are very model dependent: whether the spectrum becomes harder or softer depends on the helicity of waves, which is hard to predict, especially in a non-linear regime such as the one that is expected close the shock front. In \cite{dam,ptuskin2010} the authors assume that wave velocity can be estimated as the Alfv\'en speed as calculated in the amplified field, $v_{W} = \delta B/\sqrt{4\pi \rho}$, with $\delta B\gg B_{0}$, $B_{0}$ being the pre-existing magnetic field. 
The situation might be appreciably more complex than that: as discussed in \cite{zweibel,achterberg,blasi09} the growth rate of the resonant modes excited by CRs in the regime of high CR acceleration efficiency have a phase velocity $v_{\phi} = \left( \frac{n_{CR}}{n_{i}}v_{s}c\right)^{1/2}$, where $n_{CR}$ is the density of accelerated particles, $n_{i}$ is the ion density, $v_{s}$ the shock velocity and $c$ the speed of light. It is easy to see that $v_{\phi}$ may easily exceed the Alfv\'en speed. In this regime the growth rate of the waves also changes.

\begin{figure}
\begin{center}
\resizebox{1\columnwidth}{!}{%
  \includegraphics{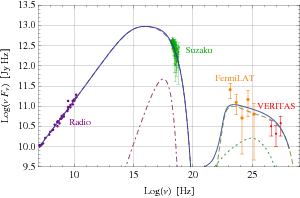} 
}
\caption{Spatially integrated spectral energy distribution of the Tycho SNR \cite{morlino}. The curves show the calculated multifrequency spectrum from the radio to the gamma ray band.}
\label{fig:tycho}       
\end{center}
\end{figure}

The simple recipe of \cite{damiano,ptuskin2010} was used to calculate the spectrum of CRs at the Earth and it was shown that the required Galactic diffusion coefficient is $D(E)\propto E^{0.54}$ (see also \cite{usnuclei}), which alleviates but does not solve the anisotropy problem. The finite velocity of the scattering centers was also used to calculate the multifrequency spectrum of the Tycho SNR \cite{morlino}. The results of this calculation are illustrated in Fig. \ref{fig:tycho} where I show the multifrequency spectrum from the radio band to gamma rays. The relatively steep spectrum in the gamma ray range, that provides a good fit to the data points from Fermi-LAT \cite{fermi} and Veritas \cite{veritas}, illustrates well the effect of the velocity of scattering centers (the gamma ray spectrum $\nu F_{\nu}$ would be roughly flat in the absence of this effect).
The predicted strength of the magnetic field reproduces the brightness profile of the non-thermal X-ray emission, as well as the synchrotron spectrum from radio to X-rays. The case of Tycho represents the first convincing instance of a SNR accelerating CRs to energies of the order to $\gtrsim 500$ TeV. 

It is worth noticing that the magnetic field amplification as observed in the form of narrow X-ray rims in SNRs might also be the result of purely hydrodynamical processes \cite{jokipii2007}, as due to the presence of density inhomogeneities upstream that result in vorticity at the shock crossing and magnetic field amplification due to the wrapping of the eddies downstream. This mechanism operates downstream of the shock and no magnetic field amplification should be expected upstream. This is a very important point in that DSA requires effective scattering of the particles on both sides of the shock and in fact if magnetic field amplification only takes place downstream, no significant increase in the maximum energy should be expected. One noticeable exception to this statement appears if the shock is quasi-perpendicular in which case particle acceleration may proceed through drifts and may in principle be very fast. In this perspective the case of Tycho is especially important, since it is a supernova of Type Ia that exploded in the normal ISM and preferential acceleration at perpendicular shocks should result in a bilateral structure that is quite unlike the spherical appearance of the Tycho SNR. 

\section{Propagation of CRs in the Galaxy}
\label{sec:propa}

The propagation of CRs in the Galaxy is best understood in terms of diffusion in a disordered magnetic field, as shown by the relative abundance of secondary nuclei and their parent primary nuclei (most notably boron versus carbon), which indicate confinement times in the Galaxy volume which exceed any ballistic time by orders of magnitude. Similar conclusions can be reached based on the abundances of some unstable isotopes (such as $^{10}Be$). The propagation of nuclei and leptons in the Galaxy is usually modeled with the help of numerical codes such as GALPROP and DRAGON. Although a substantial agreement is claimed between predictions and data, usually the price to pay in these models is that of imposing breaks in the diffusion coefficient and/or the injection spectrum in order to reproduce observations. Moreover there is some degeneracy between different models such as models with reacceleration and $D(E)\propto E^{1/3}$ and models without reacceleration and $D(E)\propto E^{0.5}$. The physical origin of these alleged breaks is unknown, which makes the fit of the data somewhat unsatisfactory from the physical point of view. The recent data by PAMELA and CREAM have made the situation more puzzling. 

\begin{figure}
\begin{center}
\resizebox{1\columnwidth}{!}{%
  \includegraphics{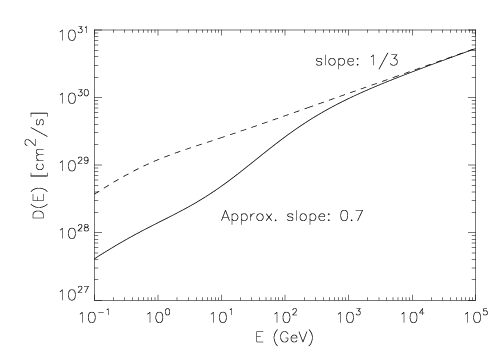} 
}
\caption{Diffusion coefÞcient induced by streaming instability of CRs and cascading from a large spatial scale of 50 pc (figure from Ref. \cite{paperI}).
}
\label{fig:diff}       
\end{center}
\end{figure}

The PAMELA \cite{pam} experiment has provided evidence that the spectra of CR protons and helium nuclei have a break at rigidity $\sim 200$ GV. The spectrum of protons measured by PAMELA has a slope $2.85\pm 0.015 (stat) \pm 0.004(syst)$ at $R<240$ GV and $2.67\pm 0.03\pm 0.05$ at $R>240$ GV. The spectrum of He nuclei has slopes $2.766\pm 0.01\pm 0.027$ and $2.477\pm 0.06\pm 0.03$ in the same rigidity ranges. The CREAM experiment measured these spectra at higher energies and found slopes compatible with those of PAMELA for $R>200$ GV (for protons the high energy slope is $2.66\pm 0.02$ and for He the slope is $2.58\pm 0.02$). This finding suggests that some new physical phenomenon might be occurring at a scale of $\sim 200$ GV either in the physics of acceleration or in the diffusion of CRs throughout the ISM. A few models have been proposed to explain the spectral hardening: \cite{malkov} suggested that protons and He may be injected with slightly different spectra as a result of a different scaling of their injection rates during the temporal evolution of the SNR. This idea is very important to understand why the two spectra are slightly different but does not address the issue of the spectral hardening observed for both p and He nuclei at $\sim 200$ GV. In \cite{tomassetti}, the author points out that a spectral hardening may be understood if the diffusion coefficient in the halo is different from that in the disk. In \cite{serpico} the spectral hardening was attributed to the interplay between the self-generated turbulence induced by CR streaming in the ISM and the non linear Landau damping that induces the cascade of turbulence from a large scale injection, possibly due to SN explosions. The problem of diffusion of CRs in this partially self-generated turbulence is non linear and can be solved only by coupling the transport equation with the equation describing magnetic field generation and damping. A transition in the diffusion properties of CRs is predicted by \cite{serpico} to appear at rigidity $\sim 200$ GV: the diffusion coefficient at low energy has an approximate scaling $D(R)\sim E^{0.7}$, while at $R>200$ GV, the scaling becomes $D(R)\sim E^{1/3}$ for a Kolmogorov phenomenology (see Fig. \ref{fig:diff}). At energies below $\sim 10$ GeV propagation becomes dominated by advection of CRs with waves moving at Alfv\'en speed away from the disc of the Galaxy. Since the self-generated waves are produced along the CR gradient, there are no waves moving towards the disc, hence one should not expect appreciable reacceleration in this model. 
The calculation of \cite{serpico} was also used to show that this approach can accomodate the spectrum of CRs inferred by \cite{neronov,kachel} using the gamma ray spectrum observed by Fermi-LAT from the direction of clouds in the Gould belt. The calculations of propagation of other nuclei including the effects of spallation of nuclei and wave generation and absorption from all nuclei is currently being done (Aloisio and Blasi, in preparation). 

The main lesson to be learnt from this model is that the ability of CRs to generate the conditions for their diffusion is a widespread phenomenon, and it is likely to be very important in the acceleration sites as well as in the propagation of CRs in the ISM. 

\section{DSA in the presence of neutrals}
\label{sec:neutrals}

It is important to realize that SNR shocks are collisionless, namely their formation is not due to particle-particle scattering but rather to the mediation of electromagnetic instabilities. The thickness of the shock front is expected to be of the order of the Larmor radius of thermal protons behind the shock. Interestingly, even electrons and protons, due to their different mass are expected to thermalize to different temperatures, $T_{e}/T_{p}\sim m_{e}/m_{p}$, behind the shock. Other collisional and collisionless processes may equilibrate electrons and protons downstream. Neutral atoms that cross a collisionless shock do not experience any jump but are coupled to the background plasma through the processes of charge exchange and ionization \cite{cheva1,cheva2}.

The presence of partially ionized material in the acceleration region may profoundly change the way DSA works, mainly because of 1) ion-neutral damping of waves which may stop the growth of CRs induced waves and hamper the acceleration process, and because of 2) the dynamical reaction of neutral material in proximity of a collisionless shock front. In addition, the presence of neutral atoms may provide us with a precious diagnostic tool of the acceleration process, as we discuss below. 

The dynamical reaction of neutrals on the shock is mainly due to the phenomenon of the neutral return flux (NRF) first investigated in the context of SNRs by \cite{paperI}. The main coupling between neutrals and ions at a collisionless shock is due to charge exchange and ionization, that are activated when the net relative speed between the two components is non-zero. Downstream of a collisionless shock ions are slowed down and heated up, while neutrals cross the shock and keep their initial velocity. The charge exchange reactions occurring in this situation may eventually produce neutrals moving with large bulk velocity in the direction of the shock and these ions may recross the shock toward upstream. Charge exchange and ionization with the upstream plasma lead to deposition of energy and momentum of these neutrals upstream, which in turn results in heating of this gas. For shock speed $\leq 4000$ km/s this NRF considerably changes the structure of the shock and leads to the formation of a shock precursor, similar to the one induced by CRs but in general on a smaller spatial scale. In \cite{paperI} the authors showed that the spectrum of test particles accelerated at such shock may visibly deviate from the standard predictions of DSA and account for steeper spectra of accelerated particles. For shocks faster than $\sim 4000$ km/s, on average a neutral crossing the shock gets ionized before suffering a charge exchange, therefore the NRF is suppressed. 
\begin{figure}
\begin{center}
\resizebox{1\columnwidth}{!}{%
  \includegraphics{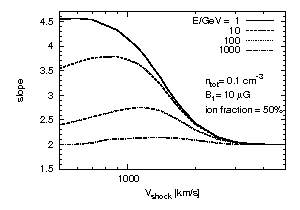} 
}
\caption{Slope of the spectrum of accelerated test particles for E = 1, 10, 100, 1000 GeV, as a function of the shock speed (figure from Ref. \cite{paperI}).
}
\label{fig:slope}       
\end{center}
\end{figure}
The net effect of the NRF is to heat the plasma in the precursor so that the Mach number of the shock decreases, namely the compression factor gets smaller. Since the spectrum of particles accelerated at the shock depends on the compression factor, the NRF leads to steeper particle spectra. Clearly the effect is the largest for particles diffusing on spatial scales comparable with the scale of the neutral induced precursor. Since the diffusion coefficient is a growing function of energy, the effect of the NRF shows more prominently at lower energies. In Fig. \ref{fig:slope} we show the slope of the spectrum of accelerated test particles as a function of the shock velocity for different values of the particle energy. The ionization fraction is assumed to be $50\%$. As expected, the spectral steepening is mostly present for $v_{sh}<3000$ km/s, and it extends to particle energies of $\sim TeV$. At high enough energy the standard slope $\gamma=2$ is recovered. 

As mentioned above, neutral hydrogen atoms in the acceleration region also provide us with an important diagnostics of the acceleration process, through their Balmer emission. The Balmer line emission from hydrogen in the shock region is becoming a powerful tool to measure the CR acceleration efficiency in SNR shocks. The idea is relatively simple: it is well known that the Balmer line produced by neutrals that suffered charge exchange with hot ions downstream has a width that reflects the temperature of ions, while neutrals that did not suffer charge exchange emit a narrow Balmer line with width $\sim 20$ km/s, corresponding to the upstream $T\sim 10^{4}$ K hydrogen temperature. If CR acceleration is efficient, part of the ram pressure upstream of the shock $\rho u^{2}$ is channelled into accelerated particles instead of heating. Therefore the plasma temperature downstream is lower if particle acceleration is efficient and the corresponding broad Balmer line becomes correspondingly narrower. On the other hand, efficient CR acceleration produces a shock precursor upstream that leads the ionized plasma to slow down with respect to neutrals. Therefore charge exchange is also activated upstream and the narrow component of the Balmer line may become broader. As shown in \cite{paperII}, for slow enough shocks the NRF also produces a similar effect but it results in the formation of a component of the Balmer line with width intermediate between the broad and the narrow line. 

The appropriate use of Balmer lines as a diagnostic tool of CR acceleration must be based on a careful modeling of the interplay between ions, neutrals and CR. A theory of NLDSA in the presence of neutrals is currently being developed. A semi-quantitative assessment of the CR acceleration efficiency in selected SNRs has been made in the past few years: in \cite{rcw86} the authors measured the width of the broad Balmer line in a region of the SNR RCW86 and found that the inferred ion temperature could be interpreted as the result of $\geq 50\%$ of the ram pressure being converted to accelerated particles. A review of Balmer lines as diagnostics of CR acceleration can be found in \cite{heng}.

\section{Implications of the SNR paradigm for the transition to extragalactic CR}
\label{sec:trans}

The amplification of magnetic field in proximity of a SNR shock is crucial to accelerate CRs up to rigidity $R\sim 10^{6}-10^{7}$ GV \cite{dam2007}. In fact higher energies are very hard to achieve unless some very special conditions are realized (fast quasi perpendicular shocks), as speculated in \cite{ptuskin2010} in connection with SN type Ic. In the minimal version of the SNR paradigm, the all-particle spectrum of CRs can be explained in terms of particle acceleration in core collapse supernovae and Type Ia supernovae: the knee results from a change in chemical composition and the Galactic CR spectrum ends with an iron dominated composition at $E\sim 2\times 10^{17}$ eV. At this point the transition to extragalactic CRs should be realized. There are basically two models of extragalactic CRs in which the transition occurs in this energy region. 

For long time it has been taken for granted that the ankle in the CR spectrum, at $\sim 10^{19}$ eV, is the spectral signature of the transition from a steep Galactic spectrum to a flatter extragalactic spectrum. The situation has however changed and the nature of the ankle questioned as a consequence of two developments: a) in Ref. \cite{dip} the authors noticed that Bethe-Heitler pair production leaves a distinct feature in the spectrum of CRs propagating on cosmological scales. The feature takes the form of a dip whose shape fits very well the observed modification factor for all experiments, with the possible exception of the one measured by the Pierre Auger Observatory. In this model CRs with energy $\ge 1$ EeV are of extragalactic origin, and the transition occurs at the second knee. b) In Ref. \cite{mix} the authors discussed the possibility that UHECRs may be nuclei with a mixed chemical composition. In this model the Galactic component of CRs ends at energy $\sim 2$ EeV. The so-called disappointing model introduced in Ref. \cite{disapp} is a special case of the mixed composition model, in which the maximum energy of protons is relatively low, $\sim 4\times 10^{18}$ eV, and the iron spectrum extends to $\sim 10^{20}$ eV. The model is {\it disappointing} in that the flux suppression at $\sim 10^{20}$ eV is not the GZK feature but rather the intrinsic cutoff in the source spectrum and no correlation with sources is expected because of the heavy composition at the highest energies. 

\begin{figure}[t]
\begin{center}
\resizebox{1\columnwidth}{!}{%
  \includegraphics{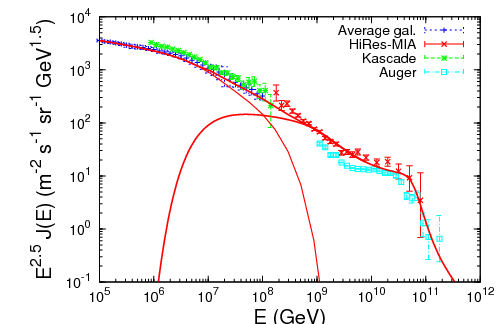} 
  }
\caption{Spectrum of CRs in the dip model overlapped to the Galactic CR flux.}
\label{fig:spectradip}       
\end{center}
\end{figure}
\begin{figure}

\begin{center}
\resizebox{1\columnwidth}{!}{%
  \includegraphics{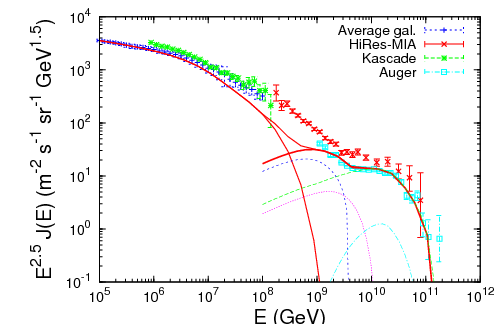} 
  }
\caption{Spectrum of CRs in the disappointing model with the Galactic CR flux as in Fig. \ref{fig:spectradip}.}
\label{fig:spectramix}       
\end{center}
\end{figure}

Both these models (dip and mixed composition) lead us to expect that Galactic CRs may end in the $<$ EeV region (with a composition dominated by heavy nuclei) rather than at the ankle. This conclusion also appears to be supported by a recent investigation of cosmic ray anisotropies \cite{giacinti}. 

The spectrum of cosmic rays is shown in Fig. \ref{fig:spectradip} for the dip model and in Fig. \ref{fig:spectramix} for the mixed composition model in its disappointing model configuration. In these calculations the spectrum of Galactic CRs was taken from \cite{amato1}. In the right panel, the different lines represent the fluxes of different chemicals, the solid line representing the total flux. One can clearly see that at high energy the chemical composition is dominated by heavy nuclei. The transition region is better described by the dip model. 

The main discrimination among dip model, mixed composition model and ankle model is based upon the measurement of the chemical composition, especially in the transition region \cite{uscomp}. In the dip model the elongation rate is expected to show a sharp transition from a heavy Galactic composition at energy below the second knee to a light composition at $E>10^{18}$ eV (see \cite{usdip}). The transition is predicted to be concluded at $10^{18}$ eV, where the composition is completely dominated by protons. The dip model works well provided the abundance of helium in the primaries is smaller than $\sim 10\%$ in flux of extragalactic CR. 

In the mixed composition model the extragalactic CRs are made of a mix of different elements, and the flux at Earth is the result of a complex chain of interactions: photodisintegration of nuclei leads to lighter composition even when starting with pure iron at the source. One can change the injection spectra and the composition at the source to match the observed composition and spectra, although in general it is hard to explain a transition to heavy composition at ultra high energies, as it is observed by Auger. In the case of the mixed composition model the elongation rate shows a gradual transition from heavy galactic CRs to somewhat lighter extragalactic CRs. The so-called disappointing model reproduces the heavy composition of Auger at high energy by construction. 

In the traditional ankle model the transition reflects in a gradual change from an iron dominated composition of Galactic CRs that extend to $\sim 10^{19}$ eV to a pure proton composition, reached at energies $\sim 5\times 10^{19}$ eV, so that in this case only a small region of energies is filled by the extragalactic CR component. This model does not fit the elongation rate as observed in any of the current experiments and is not immediately compatible with the SNR paradigm for the origin of Galactic CRs. 

\section{Summary}
\label{sec:summary}

Recent gamma and X-ray observations of SNRs have strengthened the case in favor of SNRs as the main sources of Galactic CR. However several loose ends remain as they arise when a serious attempt is made to make sense of what we measure. On one hand the spectrum of particles generated by DSA for strong shocks is very close to $E^{-2}$, this prediction is at odds with the measurement of the large scale anisotropy \cite{amato2} since it requires the diffusion to be exceedingly fast. Moreover the spectra of SNRs observed in gamma rays, when interpreted as a result of production and decay of neutral pions, suggest steeper spectra \cite{damiano}. The difference between instantaneous and escaping CR spectrum \cite{escape1,escape2} does not appear to be sufficient to explain this discrepancy. The usage of NLDSA instead of test particle DSA makes the problem even more severe, although it has been proposed that somewhat steeper spectra might be produced if the finite velocity of the scattering centers is taken into account \cite{damiano}. This idea has been tested versus the multifrequency modeling of the Tycho SNR \cite{morlino} and versus the spectra of propagated nuclei in the ISM \cite{usnuclei,ptuskin2010}. In the latter case a Galactic diffusion coefficient $D(E)\propto E^{0.54}$ was required (instead of $D(E)\propto E^{0.75}$ found by \cite{berevolk07} in the absence of this effect). 

An independent support to the SNR paradigm is provided by the detection of narrow non-thermal X-ray rims in numerous SNRs. The thickness of these filaments can be best explained in terms of magnetic field amplification, possibly induced by the streaming instability excited by CRs at SNR shocks. The observed fields are of order of $100-1000\mu G$ and may explain the acceleration of CRs up to $10^{6}-10^{7}$ GV rigidities, as required by CR data. These findings outline a situation in which the Galactic CR spectrum ends at energies $\sim (2-3) \times 10^{17}$ eV with an iron dominated chemical composition \cite{amato1}.

The simple picture that has survived for several decades of a simple diffusive propagation of CRs in the Galaxy has recently been threatened by some observational findings. The PAMELA experiment found that both the proton and He spectra show a break at rigidity $\sim 200$ GV. The He spectrum is also found to be systematically harder than the proton spectrum at all energies. The hardening of spectra might be the signature of a physical process that affects either acceleration or propagation in the ISM. In \cite{serpico} the authors suggest that this change of slope is the result of a transition from self-generated waves, where the scattering centers responsible for CR diffusion are generated by the same CRs through streaming instability, to diffusion in a background of waves cascading from larger scales for instance through non-linear Landau damping (wave-wave coupling). In this model the diffusion coefficient is an output of the calculations, and shows a complex structure, as illustrated in Fig. \ref{fig:diff}. The spectrum of CRs fits the slopes measured by PAMELA and the transition energy between the two regimes of diffusion is naturally found to be around $\sim 200$ GeV (for protons). Interestingly the slope of the PAMELA spectrum at $R<200$ GV is the same as that found by \cite{neronov,kachel} by analyzing the gamma ray spectra of clouds in the Gould belt and inferring the parent CR spectrum. The slope of the CR spectrum at $R>200$ GV is confirmed by measurements carried out with the CREAM experiment. 

Although there is no doubt that gamma ray observations are playing a crucial role in establishing the role of SNRs as the main sources of Galactic CR, it is also clear that uncertainties in the environmental conditions in which acceleration takes place creates some ambiguity in the conclusions and often these uncertainties do not allow us to discriminate between a leptonic and a hadronic origin of the observed gamma radiation. In some SNRs different regions may radiate gamma rays due to different mechanisms and the angular resolution of the gamma ray telescopes may not be sufficient to make these situations clear. Other methods to establish whether a SNR is accelerating CRs efficiently should be sought after.  In this perspective, the most interesting method is the observation of optical Balmer lines. The width of Balmer lines is affected by the presence of CRs in the acceleration region: the broad component is narrower if an appreciable fraction of the ram pressure entering the shock is being channelled into CR acceleration. On the other hand if CRs are being accelerated effectively, a precursor is formed upstream of the shock, and this may lead to a broader narrow component of the Balmer line. Spatially resolved observations of the shape of the Balmer line in SNRs may turn out to provide a wealth of new information about the process of particle acceleration in SNRs. 

\section*{Acknowledgments}
The author is grateful to the Organizing Committee of the Scineghe 2012 Workshop for their support and warm hospitality in Lecce. 







\end{document}